\documentclass[preprint]{aastex}

\usepackage{graphicx}
\usepackage{natbib}
\usepackage{pdflscape} 
\usepackage{amssymb}
\usepackage{ulem} 
\usepackage{color} 

\newcommand{\be}{\begin{equation}}
\newcommand{\ee}{\end{equation}}
\newcommand{\nn}{\mbox{} \nonumber \\ \mbox{} }
\newcommand{\ba}{\begin{eqnarray}}
\newcommand{\ea}{\end{eqnarray}}

\newcommand\eg{{\it{{e.g.,\ }}}}

\newcommand{\Lf}{{Lorentz factor}}
\newcommand{\Bf}{{magnetic field}}

\newcommand{\NS}{neutron star}

\newcommand{\BH}{{black hole}}

\begin{document}

\title{Ultra-relativistic double explosions}

\author{Maxim Lyutikov\\
Department of Physics, Purdue University, \\
 525 Northwestern Avenue,
West Lafayette, IN
47907-2036 }

\begin{abstract}
We consider   fluid dynamics of  relativistic double explosion - when a point explosion with energy $E_1$ is followed by a second explosion with energy $E_2$ after time $t_d$ (the second explosion could be in  a form of   a long lasting wind). The primary explosion creates a self-similar relativistic blast wave propagating with \Lf\ $\Gamma_1(t)$. A sufficiently strong second explosion, with total energy  $E_2 \geq 10^{-2} E_1$, creates a fast second shock in the  external fluid previously  shocked by the primary shock. 
At times longer than the interval between the explosions $t_d$, yet short compared with the time when the second shock catches up the primary shock at  $\sim t_d \Gamma_1^2$, the structure of the second shock is approximately self-similar. Self-similar structure of the second shock exist  for the case of constant  external density (in this case $\Gamma_2 \propto t^{-7/3}$),  but not for the  wind environment.
  At early times the \Lf\ of the second shock may exceed that of the primary shock and may boost  the synchrotron emission of locally accelerated electrons into the Fermi LAT range.
 \end{abstract}

\maketitle
\section{Introduction}

We are confident  that Gamma Ray Bursts  (GRBs) are produced in relativistic explosions \citep{Paczynski86,piran_04}.  The standard fireball model \citep{piran_04} postulates that  (i) the prompt emission is produced within the ejecta; (ii) afterglows are generated in the relativistic blast wave after the ejecta deposited much of its bulk energy into circumburst  medium. This was a well-established paradigm before the launch of {\it Swift} mission.

 {\it Swift} satellites allowed us to probe early afterglows, on time scales shorter than $\sim $ a day \citep{2016ApJ...829....7L}. Most surprising are the highly-variable early $X$-ray and optical afterglows, that challenge the   standard fireball model \cite{2010ApJ...720.1513K}.  The launching mechanism, the nature of the central source, the composition of the primary wind, prompt radiation mechanisms remain uncertain \citep{lyutikov_09}

One of the  key unexpected conclusions that emerged  as a result  of the 
observations of early GRB afterglows is  that the central source could remain  active (keeps producing relativistic wind) for long times  after the explosion, $ \sim 10^3-10^4$ seconds, and perhaps even longer \citep{2007ApJ...665..599T}. 
This motivates us to consider the  following hydrodynamic problem: an initial powerful explosion generates an ultra-relativistic shock wave. It is then followed by a second explosion or a wind, which is could be {\it  subdominant energetically}. What is the structure of  outflows generated by such double explosion?

\section{A range of double self-similar outflows}

Explosions and/or winds with power-law scaling of power with time produce self-similar structures of the flow behind a forward shock \cite{Sedov,BlandfordMcKee}; magnetized forward shocks have been considered by \cite{2002PhFl...14..963L}. Let's  assume that the primary  explosion was an instantaneous event. It create a self-similar post-primary shock flow described by the Blandford-McKee, \cite{BlandfordMcKee},   solution (B\&Mc below). If there is a second  explosions (which actually can be in a form of a long-lasting wind) after some time $t_d$, the structure is not generally self-similar - since now there is a special point in time $t_d$ and, in addition, there is a special parameter -  the ratio of the energies  of the  primary and second explosions, $E_1/E_2$. But - this is the key point for the current approach - at times much smaller than the time it takes for the second shock wave to catch with the primary one (or long afterward) the flows are {\it approximately}  self-similar. At very early times the second shock propagates through a velocity and pressure gradients created by the primary shock; importantly, as long as the primary shock is in self-similar stage, those velocity and pressure gradients are self-similar, power-law like. Thus, at early times the second shock propagates in a time and coordinate scale-free environment, determined by the primary shock \citep[for the non-relativistic case, see][]{1981JAMTP..22..545A}.

In the case of GRBs the observer time $t_w$ of early afterglow variability  corresponds to physical (coordinate) time after the explosion which is much longer, $t \sim t_{w} \Gamma^2 \sim $ months.
On the other hand the observed  variability time $t_w$ is of the same order as the intrinsic variability of the source. 
Thus, in the case of  GRBs with long-lasting central source we typically have a separation of temporal scales: when the early afterglow emission is produced we have $t_w \ll  t \leq t_{w} \Gamma^2$. We expect that in the time range the double-explosion structure will be self-similar.   

The long-lasting wind from the central source  is most likely to be  highly relativistic, with \Lf\ much larger than the  \Lf\  of the primary shock. (The models of the long lasting engine include, for  example,  formation of a long-lasting \NS,  \cite{2011MNRAS.413.2031M},  or a black hole that can retain \Bf\ for times much longer than predicted by the no hair theorem,  \cite{2011PhRvD..84h4019L,2013ApJ...768...63L}.)  Also, the wind propagates through a cavity cleared by the primary shock, so it does not decelerate until it starts interacting with the material swept-up by the primary shock. 

\begin{figure}[h!]
 \centering
 \includegraphics[width=.99\columnwidth]{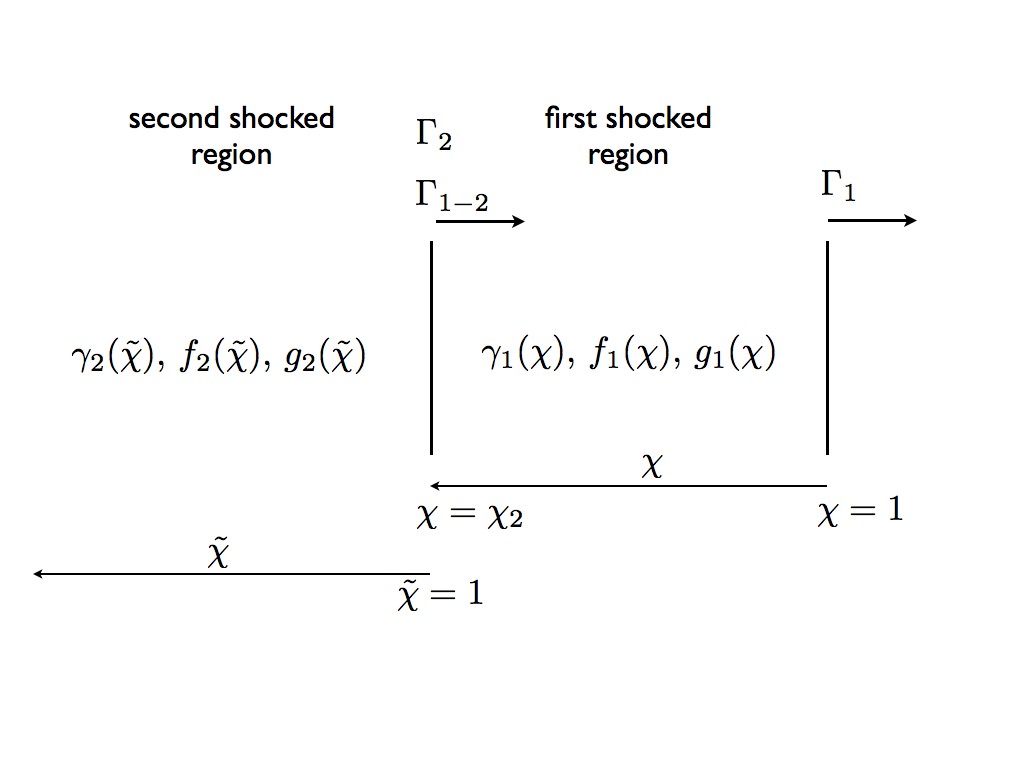}
 \caption{Cartoon of the problem. The primary shock propagates with \Lf\ $\Gamma_1(t)$. The post-primary shock flow ($\gamma_1,\, p_1 $ and $n_1$) is self-similar, described by the self-similar coordinate $\chi$ of B\&Mc. The second shock propagates with $\Gamma_2(t)$ and is located at $\chi=\chi_2 (t)$. Its instantaneous \Lf\ with respect to the shocked flow just in front of it is $\Gamma_{1-2}$. The flow parameters behind the second shock ($\gamma_2,\, p_2 $ and $n_2$) can be described by the self-similar parameter $\tilde{\chi}$.
}
 \label{Shock-structure001}
\end{figure}

When the high-\Lf\ secondary wind will catches up with the flow generated by the primary shock wave it will launch a second shock in the swept-up material, Fig. (\ref{Shock-structure001}). 
The flow between the shocks (the ``first shock region'' in Fig.  (\ref{Shock-structure001}))  is self-similar, described by B\&Mc solution with  the self-similar parameter $\chi$. 
As we demonstrate, the structure of the flow past the second shock can be {\it approximately} self-similar.

\section{Double explosion in constant density environment}

\subsection{Governing equations}

Assuming a spherically symmetric outflow, the relativistic fluid equations 
  read \cite{LLVI}
\ba 
&&
\partial_t \left[  w\gamma^2 -p\right]+
{1\over r^2} \partial_r \left[ r^2  w \beta \gamma^2 \right] =0
\label{x4} \\ &&
\partial_t \left[  w \gamma^2 \beta \right]+
{1\over r^2} \partial_r 
\left[r^2\left(w \beta^2 \gamma^2 + p\right) \right] - 
{2 p \over r}=0
\label{x51} \\ &&
\partial_t \left[ \rho \gamma \right]+
{1\over r^2 } \partial_r  \left[r^2 \rho  \beta \gamma \right] =0
\label{x6}
\ea
where all notations are standard. (We choose to  work consistently with proper quantities, 
i.e. measured in the plasma
rest frame. One should be careful in comparing our equation with B\&M below.)

Consider relativistic point explosion of energy $E_1$ in a medium with constant density $\rho_{ex}=m_p n_{ex}$,  followed by another explosion after time $t_d$. The second explosion can be a wind.
The second shock propagates through relativistically hot and bulk-moving plasma with the self-similar parameters created by the primary explosion (B\&Mc)
\ba &&
\Gamma_1 = \sqrt{ \frac{17}{8 \pi }}   
\sqrt{ \frac{E_1}{\rho_{ex} c^5}}  t^{-3/2}
\nn &&
p_1= \frac{2}{3} \rho_{ex} c^2 \Gamma_1 ^2 f_1(\chi)
\nn &&
\gamma^2_1 = \frac{1}{2}\Gamma_1 ^2 g_1(\chi)
\nn &&
n_1= 2 n_{ex} \Gamma_1 n_1(\chi)
\nn &&
f_1(\chi)=\chi^{-17/12}, 
\nn && 
g_1(\chi) =1/\chi
\nn &&
n_1(\chi)= \chi^{-5/4}
\ea
(subscript $1$ indicates that quantities are measured between the two shocks).

Suppose that the second explosion occurs at  time $t_d$ after the initial one and the second shock  is moving with the  \Lf\ $\Gamma_2 ^2 \propto (t-t_d)^{-m} \gg \Gamma_1 ^2$.  Then, the location of the second shock at time $t$ is 
\be
R_2 = (t-t_d) \left( 1-\frac{1}{2 \Gamma _2^2 (m+1)} \right)
\ee
The corresponding self-similar coordinate of the second shock in terms of the primary shock self-similar parameter $\chi$  is 
\be
\chi_2 =(1 +8 \Gamma_1^2 ) (1-\frac{R}{t}) 
\approx
\left( \frac{8 t_d}{t} + \frac{4}{(m+1) \Gamma_2^2} \right) \Gamma _1^2
\ee
For $t_d \geq t/(2 (m+1) \Gamma_2^2)$
\be
\chi_2 =
\frac{8 \Gamma _1^2  {t_d}}{t}\propto t^{-4}.
\label{chi2}
\ee
Thus, the second shock with time approaches the primary (since $\chi$ decreases with time), but  as long as $ t \leq 8  t_d \Gamma_1^2$ the second shock is located far down stream of the primary shock, at  $\chi_2 \gg 1$. Thus, the second shock does not affect the bulk of the flow created by the primary shock for a  very  long time, much longer that the time between the two explosions.
After the second shock catches up with the primary shock, at   time $\sim  t_d \Gamma^2 \gg t_d$, the  system behaves as a single explosion with the sum of energies $E_1+E_2$.  
On the other hand, for $t \geq t_d$,
transient effects associated with the details of the second explosion die out, so that for times $t_d \ll t \ll 8  t_d \Gamma^2$ the structure of the flow after the second explosion becomes self-similar. The purpose of this paper is to investigate this self-similar structure.

\subsection{Dynamics of the second shock}

Before finding the behavior of the post-second shock variables let us do an order of magnitude estimates of the temporal behavior of the second shock. 
 The total amount of enthalpy created by the first shock (enthalpy density for ultra-relativistic gas is four times pressure)  that ends up behind the second shock, at $\chi > \chi_2 $, is
\be
  w_2 = 16 \pi \int r^2 p dr= \frac{4\pi}{3} m_p c^5 n_{ex} t^3 \int_{\chi_2}^\infty \frac{d\chi}{\chi^{17/12}}=
  \frac{16 \pi ^{17/12}}{5\ 17^{5/12}} \left(\frac{c^{85} {m_p}^{17} n_{ex}^{17}}{E_1^5 t_d^5} \right) t^{14/3}
  \label{w2}
  \ee

    For a point second explosion,  the energy $E_2$ should equal this integrated enthalpy times $\Gamma_2^2$,
    \be  
    E_2\approx w_2 \Gamma_2^2
        \label {E20}
    \ee
Assuming $  \Gamma_2^2  \propto t^{-m}$ gives $m=14/3$ and the scaling for the second shock
    \be
    \Gamma_2^2 \approx
     \left(\frac{{E_1}^{5} {E_2}^{12} {t_d}^{5}}{c^{85} ({m_pn_{ex}})^{17} }\right)^{1/12}
    t^{-14/3}
   \label{Gamma2}
   \ee
Thus,
\be
\frac{\Gamma_2^2} {\Gamma_1^2} \approx \frac{E_2}{E_1} \left(  \frac{ t_d \Gamma_1^2}{t} \right)^{5/12}
 \approx \frac{E_2}{E_1} \left(  \frac{ t_d }{t_{ob}} \right)^{5/12}
\ee
where we defined the observer time associated with the first explosion, $t_{ob} \sim t/(2 \Gamma_1^2)$. 
$\Gamma_2 \propto t_{ob}^{-7/12}$.

Condition $t_d \geq t/(2 (m+1) \Gamma_2^2)$ (for Eq. (\ref{chi2}) to apply) requires 
\be
t\leq 0.47 \left(\frac{E_2}{E_1} \right)^{12/17} t_d \Gamma_{FS}^2.
\ee

This implies that as long as $t \ll t_d \Gamma_1^2$ (well before the second shock catches with the first) even energetically subdominant second explosion produces a fast second shock,
$\Gamma_2 \geq \Gamma_1$. 
(In order to calculate the coefficient in front of this relation we need to know the post-second shock flow; it is actually large $\approx 125$, see (\ref{Gamma22}) so that even energetically subdominant second explosion, with $E_2 \geq 10^{-2}E_1$ can produce fast second shocks up to the catch-up time, $\sim t_d \Gamma_1^2$.) 

We stress that the self-similar  solution behind the second shock is only approximate. For example, 
 the energy within the second shocked region increases even  for a point secondary explosion since the second shock advances in the self-similar coordinate tied to the first shock. The energy associated with the first shock, that ends up behind the second shock is
\be 
E_{1,2} \approx \frac{2 \pi}{3} t^3 \Gamma_1^2 \int_{\chi_2(t)} ^\infty \chi^{-29/12} d \chi  \propto t^{17/3}
\label{E12}
\ee
But for $\chi_2 \gg 1 $ only small fraction of the total energy resides behind the second shock; thus, we assume that the energy (\ref{E20}) is much larger than (\ref{E12})

\subsection{Self-similar solutions for the second shock}

The second shock is located in terms of the first self-similar coordinate  at $\chi=\chi_2(t)$; let us consider the  post-second shock quantities. Using strong relativistic shock conditions  \citep{LLVI},
pressure and density immediately after the second shock (assuming $\Gamma_2 \gg \gamma_1(\chi_2)$)  are 
\ba &&
p_2^{(0)} = \frac{8}{3} p_1 \Gamma_{1-2}^2 = \frac{2^{7/4}}{9} \rho_{ex} c^2 \left( \frac{t}{t_d}\right)^{5/12} \frac{\Gamma_2^2}{\Gamma_{1}^{5/6}}
\nn &&
n_2^{(0)} =2 n_1(\chi_2) \Gamma_{1-2}=\frac{\rho_{ex}}{2^{3/4}} \left( \frac{t}{t_d}\right)^{3/4}\frac{\Gamma_2}{\Gamma_{1}^{3/2}}
\nn &&
 \Gamma_{1-2}= \frac{\Gamma_2}{ 2 \gamma_1(\chi_2)}
 \ea
$  \Gamma_{1-2}$ is the relative \Lf\ of the second shock  with respect to the local plasma flow in front of it, downstream from the  first shock.

Next we introduce new self-similar variable associated with the second shock
\ba &&
{\tilde{\chi}} =( 1 + 2(1+m) \Gamma_2^2)(1-\frac{r}{t-t_d}) \approx 2(1+m) \Gamma_2^2(1-\frac{r}{t})
\ea
where 
\be
\Gamma_2 ^2 \propto (t-t_d)^{-m}  \approx  t^{-m} 
\ee
(so that ${\tilde{\chi}}=1$ at the location of the second shock),
we parametrize
\ba &&
p_2 = p_2^{(0)}  {\Gamma_2^2} f_2 ({\tilde{\chi}}), 
\nn &&
 \gamma_2^2 = \frac{1}{2} {\Gamma_2^2 g_2 ({\tilde{\chi}})}
 \nn &&
 n_2 = n_2^{(0)} \Gamma_2 n_2({\tilde{\chi}})
 \label{anzats}
 \ea

Using {\it anzats} (\ref{anzats}), expanding in $\Gamma_2 \gg 1$ and taking a limit $t\gg t_d$, we
find
\ba &&
\frac{1}{g_2}\partial _{\tilde{\chi}} \ln f_2 =-\frac{g_2 (3 m-17) {\tilde{\chi}} -24 m+44}{3 (m+1) \left(g_2^2 {\tilde{\chi}} ^2-8 g_2 {\tilde{\chi}} +4\right)}
   \nn &&
\frac{1}{g_2}   \partial _{\tilde{\chi}} \ln  g_2 = -\frac{g_2
   (m+2) {\tilde{\chi}} -7 m+9}{(m+1) \left(g_2^2 {\tilde{\chi}} ^2-8 g_2 {\tilde{\chi}} +4\right)}
   \nn &&
\frac{1}{g_2}   \partial _{\tilde{\chi}} \ln  n_2 = -\frac{g_2^2 (m-12) \chi ^2+g_2 (67-11 m) \chi +22 m-58}{2 (m+1) \left(g_2 \chi -2\right) \left(g_2^2 \chi
   ^2-8 g_2 \chi +4\right)}   \label{main}
   \ea
   with boundary conditions $g_2(1) = f _2(1)  = 1$. 
 
The point second explosion corresponds to   $m=14/3$. In this case  the post-second shock variables are
\ba && 
f_2({\tilde{\chi}})={\tilde{\chi}}^{- 71/51}
\nn &&
g_2({\tilde{\chi}})={\tilde{\chi}}^{-1}
\nn &&
n_2({\tilde{\chi}})={\tilde{\chi}}^{-53/34}
\label{sol22}
\ea
 Knowing the post-second shock solutions  (\ref{sol22}) we can calculate the coefficient in (\ref{Gamma2}) for the case of point secondary explosion, $m=14/3$. We find
  \ba &&
  \frac{\Gamma_2^2} {\Gamma_1^2} = 0.35   \frac{E_2}{E_1} \left(  \frac{ t_d \Gamma_1^2}{t} \right)^{5/12}
  \nn &&
  \Gamma_2 =\sqrt{\frac{71}{2}} \left(\frac{17}{\pi }\right)^{5/24}
\left(  \frac{{E_1}^5 {t_d}^5}{c^{85} (m_p n_{ex}) ^{17} }\right)^{1/24} \sqrt{E_2} t^{-7/3}
  \label{Gamma22}
\ea
  Thus, for $E_2 \geq 10^{-2} E_1$, there is a range for the self-similar solutions to be applicable all the way to $ t \sim \Gamma_1 ^2 t_d$ (the time when the second shock catches with the primary shock). 

\begin{figure}[h!]
 \centering
 \includegraphics[width=.49\columnwidth]{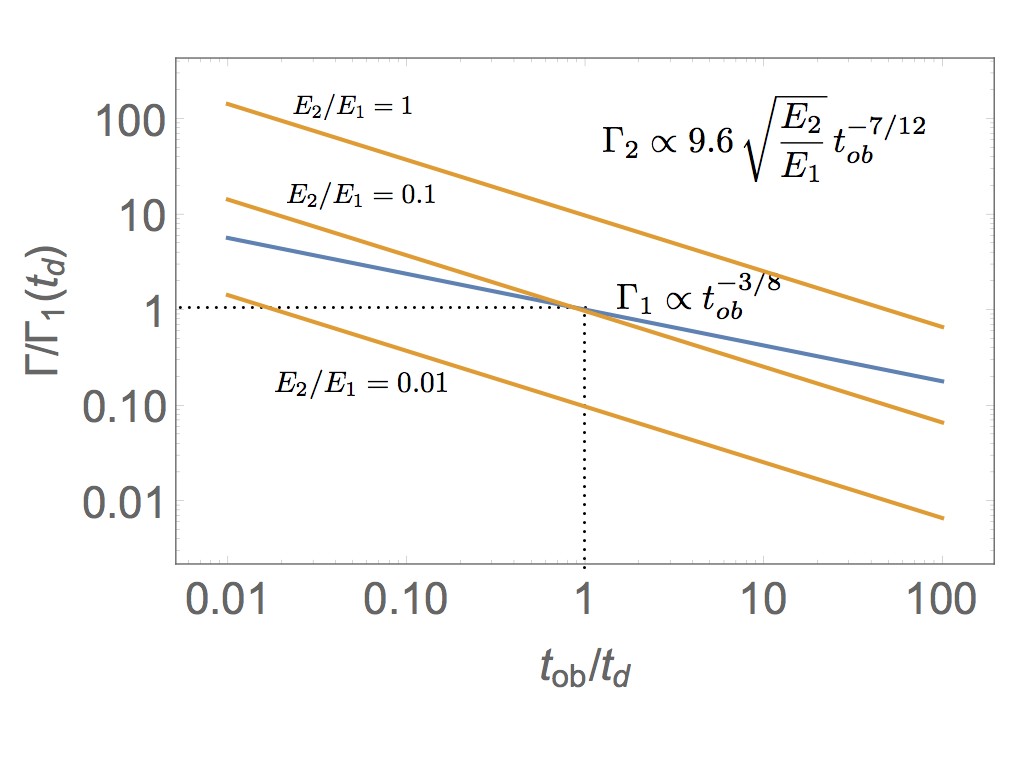}
  \includegraphics[width=.49\columnwidth]{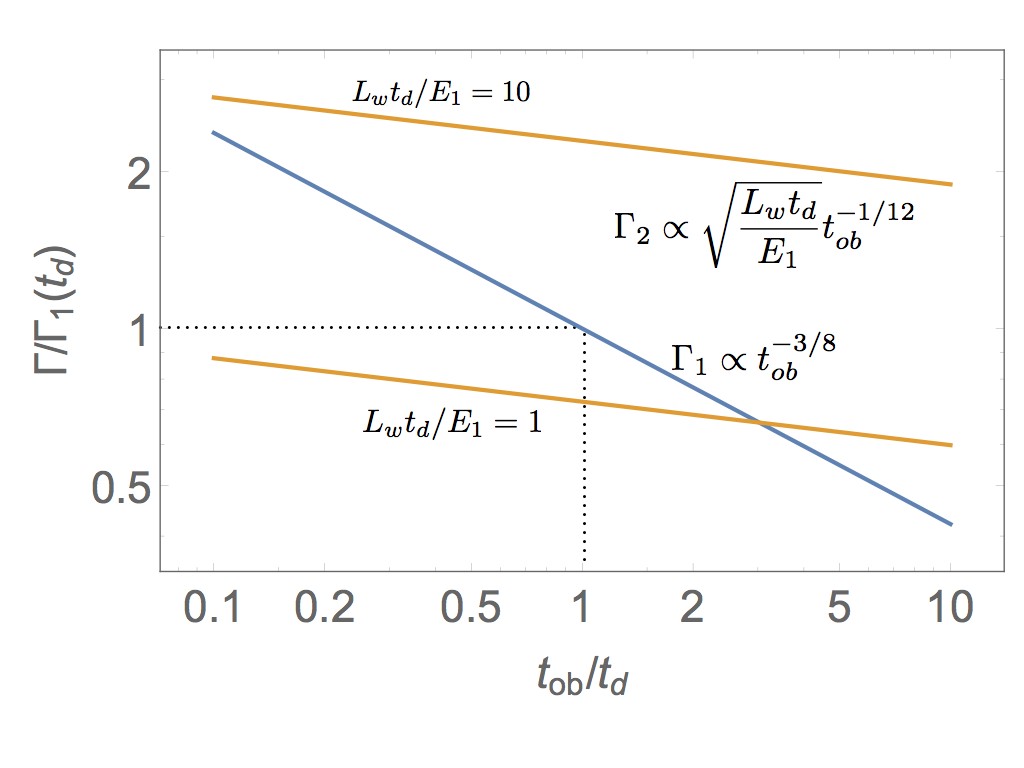}
 \caption{Evolution of the \Lf\ of the second shock as a function of observer time $t_{ob}$ associated with the primary shock, $t_{ob} = t/(2 \Gamma_1^2)$.
{\it  Left Panel}: point explosion, {\it  Right Panel}:  constant luminosity source.
Lorentz factors are normalized to $\Gamma_1(t_d)$. 
 Different curves are parametrized by the ratio of second to first  energies (for point explosions) and the total energy produced in time $t_d$ by the source (for constant luminosity case).
 Thus: (i) at times $t_{ob} \leq t_d$ the secondary shock is much faster than the primary shock even for the case of energetically subdominant second explosion; (ii) winds are ineffective in driving fast second shocks.
}
 \label{Shock-structure0031}
\end{figure}

If expressed in terms of the observer time for the first shock $t_{ob}= t/(2 \Gamma_1^2)$, 
 \ba&&
  \frac{\Gamma_2} {\Gamma_1} = 
   9.6 \, \sqrt{  \frac{E_2}{E_1} } \left(  \frac{ t_d }{t_{ob}} \right)^{5/24},
  \ea
see Fig. \ref{Shock-structure0031}.

Thus, for observer's time (defined with respect to the first explosion) shorter than the delay between the two explosions, the \Lf\ of the second may  exceed the \Lf\ of the primary shock for $E_2 \leq E_1$. What is more, for observer times shorter than the delay time, and not too weak second explosions the \Lf\ of the second shock might exceed the first short by $\sim $ an order of magnitude.

\subsection{Continuos energy injection}

For a long-lasting central source producing luminosity $L_w \propto t_e^{q}$ with $t_e \propto t/\Gamma_2^2$ the energy deposited into the post-second shock flow scales as $ \propto L_w t /\Gamma_2^2 \propto t^{(1+m)(1 +q)}$.Thus,
\be
m = \frac{11-3q}{3(2+ q)}
\ee
For constant luminosity source, $q=0$, $m=11/6$.

Qualitative estimates for the  power absorbed by the shocked medium
\be
L_0 t_{em}^q \frac{ t}{\Gamma_2^2} = w_2 \Gamma_2^2
\ee
with $t_{em}= t/\Gamma_2^2$ and $w_2$ given by (\ref{w2})
we confirm  $m=11/6$. The power balance gives
\ba &&
\frac{\Gamma_2^2}{\Gamma_1^2} \approx  \sqrt{ \frac{L_w t}{E_1}} \left( \frac{ t_d}{t \Gamma_1^2}\right)^{5/24}=
 \sqrt{ \frac{L_w t_{ob} }{E_1}} \left( \frac{ t_d}{t _{ob}2}\right)^{5/24}
 \nn &&
 \Gamma_2= \frac{\sqrt{L_w } t_d ^{5/24}}{(E_1^3 n_{ex} m_p c^5)^{1/8} t_{ob}^{1/12}}
\ea
Thus, to have  $\Gamma_2 \geq  \Gamma_1$ it is required  for the applicability of the self-similar approach that the power be sufficiently strong,
\be
L_w \geq \frac{E_1}{ t_d ^{5/12}  t_{ob}^{7/12}}
\ee
Since both $t_d$ and $t_{ob}$ are $\sim 10^4$ sec, then for the initial explosion of $10^{52}$ erg it is required that $L_w \sim 10^{48} $ erg sec$^{-1}$.)

For $m\neq 14/3$, introducing new variable $x=\tilde{\chi} g_2$, Eqns (\ref{main})
take the form
\ba &&
\partial _{x} \ln f_2 =\frac{3 m (x-8)-17 x+44}{3 \left(m (x-4)+x^2+17 x-4\right)}
   \nn &&
  \partial _{x} \ln  g_2 =\frac{m (x-7)+2 x+9}{m (x-4)+x^2+17
   x-4}  
   \nn &&
     \partial _{x} \ln  n_2=\frac{m \left(x^2-11 x+22\right)-12 x^2+67 x-58}{2 (x-2) \left(m (x-4)+x^2+17 x-4\right)}
    \label{main1}
   \ea
   with solutions
   \ba &&
   Z_1 = \frac{-m x+4 m-x^2-17 x+4}{3 m-14}
   \nn && 
   Z_2 =\frac{\left(1-\frac{m+19}{\sqrt{m^2+50 m+305}}\right) \left(1+\frac{m+2 x+17}{\sqrt{m^2+50
   m+305}}\right)}{\left(1+\frac{m+19}{\sqrt{m^2+50 m+305}}\right) \left(1-\frac{m+2 x+17}{\sqrt{m^2+50
   m+305}}\right)}
   \nn && 
   f_2 = Z_1^{\frac{1}{6} (3 m-17)} Z_2^{-\frac{-3 m^2-82 m+377}{6 \sqrt{m^2+50 m+305}}}
   \nn && 
   g_2 = Z_1^{\frac{m+2}{2}} Z_2^{\frac{m^2+33 m+16}{2 \sqrt{m^2+50 m+305}}}
   \nn &&
   n_2= (2-x)^\frac{m+7}{17-m} Z_1^\frac{m^2-27 m+218}{4 (m-17)} Z_2 ^ \frac{m^3+4 m^2-737 m+4636}{4 (m-17) \sqrt{m^2+50 m+305}}
   \label{solx}
   \ea

\begin{figure}[h!]
 \centering
 \includegraphics[width=.99\columnwidth]{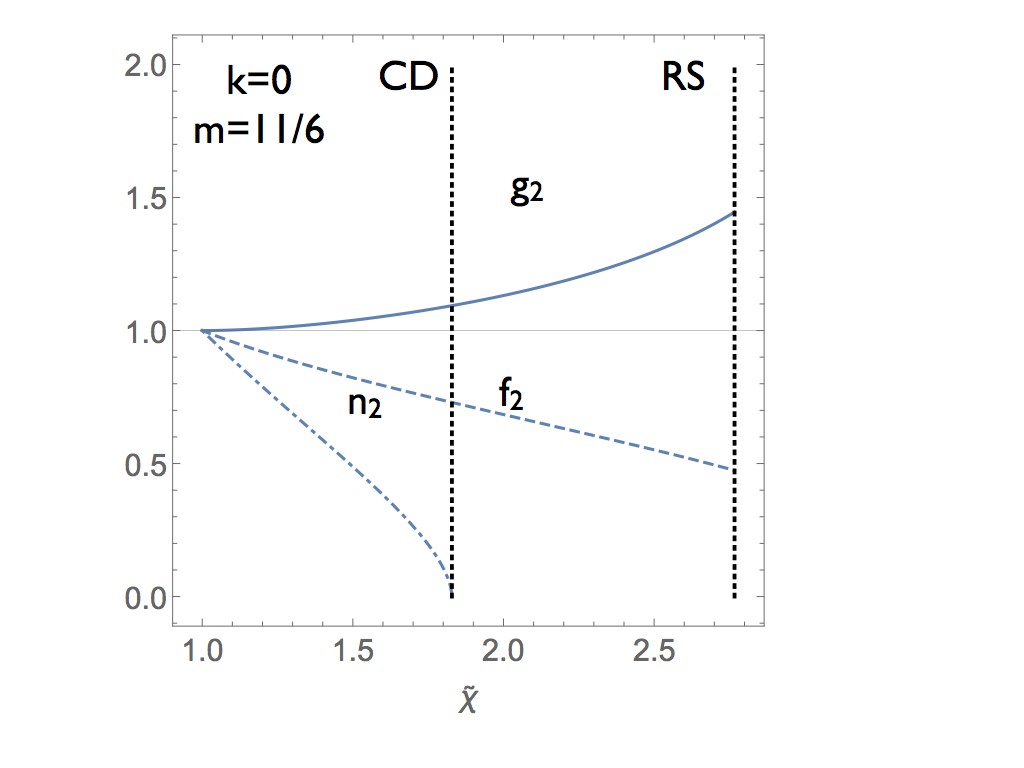}
 \caption{Structure of  the post-second shock flows for constant luminosity source following a point explosion in a constant density environment ($k=0$, $m=11/6$). Second shock is located at $\tilde{\chi}=1$. CD denotes the location of the contact discontinuity, RS - of the reverse shock (for the case of fluid wind, with zero  magnetization.)
}
 \label{Shock-structure002}
\end{figure}

The  contact discontinuity (CD) is at $x=2$ and the wind termination shock is at $x=4$. Since for the point  second explosion case $x\equiv 1$, these values are never reached in that case, naturally. 
Thus solutions (\ref{solx}) should not be extended beyond $x=4$.
For $m=11/6$,  the point $x=2$ is reached when $\chi = 1.82$, $g_2=1.09$, $f_2 =0.73$. At the termination shock $x=4, \,  \chi =2.76, \, f_2 =0.47, \, g_2 =1.44$. Density is zero on the CD, Fig. \ref{Shock-structure002}.

\section{Applications to GRBs}

\subsection{Early GeV emission}

Some GRBs show emission in the {\it Fermi} LAT detector, in the $100$ MeV - $100$ GeV range \cite{2009ApJ...706L.138A,2014Sci...343...42A}. The LAT photons start arriving during the prompt phase, and continue well after the prompt phase has ended. The earlier photons have typically lower energy, few hundred MeV, while late photons can reach energies $\sim 100$ GeV in the explosion rest frame \cite{2014Sci...343...42A}. If the mechanism of GeV photon production is synchrotron, then such high energy require both very fast rate of particle acceleration and large bulk Lorentz factors, $\geq 10^3$ \cite{2010MNRAS.405.1809L,2013arXiv1306.5978L}.

Late GeV photons are more naturally explained by the inverse Compton scattering \citep{2014ApJ...788...36B}, but the early, lower energy ones can be synchrotron from the fast second shock. For example, using a standard parametrization for the external shock, $\gamma_e \sim (m_p/m_e) \epsilon _e \Gamma_2$, $ b_2^2/(8\pi)=\sqrt{ \epsilon_B} (8/3) m_p n_{ex} c^2$, the typical frequency of the synchrotron photons in the second shock can be estimated as
\be
\epsilon_s = \hbar \gamma_e^2 \Gamma_2 \frac{e b_2} {m_e c} = 6 \times 10^{-3}
\epsilon _e^2 \sqrt{ \epsilon_B} \sqrt{n_{ex} } \Gamma_2^4=
2.5 \times 10^8 \frac{t_d^{5/6}}{t_{ob}^{7/3}} \, {\rm eV} 
\label{epsilons}
\ee
(for $\epsilon_e = \epsilon_B =0.1$).
This falls into the LAT range  for $t_d \sim t_{ob}$.

\subsection{Contribution from the second shock  to the  afterglow emission}

\begin{figure}[h!]
 \centering
 \includegraphics[width=.49\columnwidth]{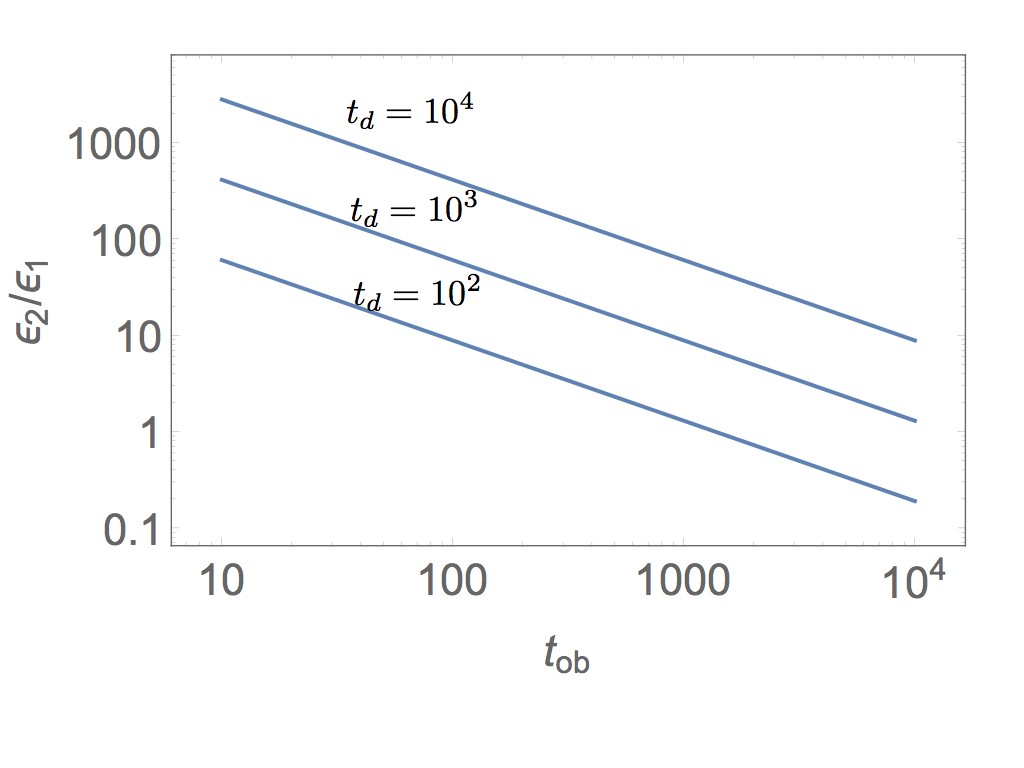}
  \includegraphics[width=.49\columnwidth]{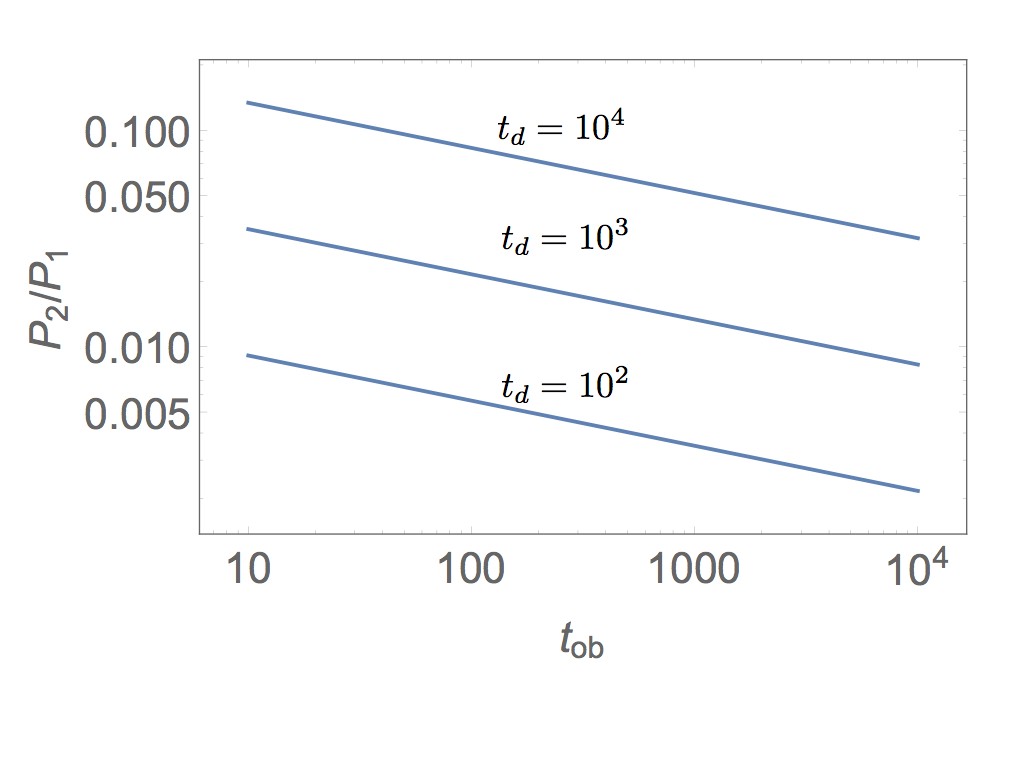}
 \caption{ Ration of synchrotron peak energies {\it  Left Panel} and peak power  {\it  Right Panel} for second and first shocks in terms of the observer time $t_{ob}$. Ratio of energies is $E_2/E_1=10^{-2}$, $E_1 =10^{52}$ erg. 
 Plotted are curves for different delay times $t_d =10^2, \, 10^3, \, 10^4$.
}
 \label{emit}
\end{figure}

The second shock also contributes to the afterglow emission. The ratio of the typical frequencies for emission from the first and second shocks are
\be
\frac{\epsilon_{s,2}}{\epsilon_{s,1}}= \left(\frac{\Gamma_2}{\Gamma_1} \right)^4
\ee
This ratio becomes $\sim 1$ later on, when the second shock swept up enough material and decelerated, Fig. \ref{emit}. 
The ratio  of the emitted powers of the primary and the secondary shock depend on the amount of the material swept-up by the second shock $M_2$. 
We find
\be
\frac{M_2}{M_1}= \frac{t_{ob}^{1/4}}{ \sqrt{2} t_d^{1/4} \Gamma_1} \ll 1
\ee
The ratio of powers (assuming similar acceleration efficiencies at the primary and second shocks)  is then
\be
\frac{P_2}{P_1}= \left(\frac{\Gamma_2}{\Gamma_1} \right)^4\frac{M_2}{M_1}
\ee
It is typically smaller than unity, Fig. \ref{emit}. Since both shocks are self-similar, the combined emission from two shocks cannot explain flares and/or plateaux.

\section{Discussion}
In this paper we considered the self-similar structure of relativistic double explosions. For time longer than the interval between the explosions but short compared with the time that the second shock catches with the first one, the structure of the flow behind the  second shock is approximately self-similar. Since the second shock propagates through a medium cleared by the primary shock, the \Lf\ of the second shock can greatly exceed that of the primary shock.

We found that  (i) at times $t_{ob} \leq t_d$  (this corresponds approximately to coordinate  times when the second shock has not caught-up yet with the primary shock)  the secondary shock is much faster than the primary shock even for the case of energetically subdominant second explosion; (ii) winds are ineffective in driving fast second shocks - only very powerful winds (that release the energy of the order of the energy of the primary explosion on the time scale of the delay between two winds) can produce fast secondary shocks.

 Importantly,  
we do not specify the nature of the secondary explosion (the initial energy content). For example,  the second explosion can be purely magnetic,  \cite{2006NJPh....8..119L}. \eg\  due to  a long lasting wind generated by the long-lived \NS\ or a \BH. Highly magnetized winds are expected to be very fast, and thus can  produce very fast second shocks.

I would like to thank Rodolfo Barniol Duran, Dimitrios Giannios and Juan Camilo Jaramillo   for discussions. 

This work had been supported by   NSF  grant AST-1306672 and DoE grant DE-SC0016369.

 \bibliographystyle{apsrev}
  \bibliography{/Users/maxim/Home/Research/BibTex}

  \end{document}